\documentclass[11pt]{article}
\usepackage[utf8]{inputenc}
\usepackage[english]{babel}
\usepackage{graphicx}
\usepackage{sidecap} 
\usepackage{amsmath}
\usepackage{upgreek}
\usepackage{amssymb}
\usepackage[unicode]{hyperref}
\usepackage[tableposition=top]{caption}
\usepackage{multirow}
\usepackage{tabularx}
\usepackage{textcomp}
\usepackage{gensymb}
\usepackage{graphics}
\usepackage{color}
\usepackage[dvipsnames]{xcolor}
\usepackage[a4paper, total={6.5in, 8.5in}]{geometry}

\numberwithin{figure}{section}
\numberwithin{table}{section}
\usepackage{chngcntr}
\counterwithout{figure}{section}
\counterwithout{table}{section}
\usepackage{soul}
\usepackage{fancyhdr}
\usepackage{tikz}
\usepackage{orcidlink}

\title{A low-cost plasma source aimed for medical applications using Ar as the working gas}

\begin{document}

\pagestyle{fancy}
\fancyhead{} 
\fancyhead[L]{\textbf{\tiny This paper has been accepted for publication in Physica Scripta. This is the author's version which has not been fully edited and content may be different from the final publication. Citation information: F. do Nascimento, B. H. da Silva Leal, and K. G. Kostov, Phys. Scr., vol. 100, no. 9, p. 095601, Sep. 2025, DOI \href{https://dx.doi.org/10.1088/1402-4896/adfe2b}{10.1088/1402-4896/adfe2b}}}

\author{Fellype do Nascimento\orcidlink{0000-0002-8641-9894}, Bruno Henrique da Silva Leal\orcidlink{0000-0003-0524-2378},\\ Konstantin Georgiev Kostov\orcidlink{0000-0002-9821-8088}\\
\small{\textit{Faculty of Engineering and Sciences in Guaratinguetá, São Paulo State University–UNESP}}}

\vspace{10pt}

\maketitle

\begin{abstract}
Due to advances in equipment and intense research on plasma biomedical applications over the past few years, plasma devices are now suitable for clinical use. It has been demonstrated that plasma sources can be designed in a way that ensures their safe operation and application whilst preserving the relevant clinical results. However, the manufacturing and operating costs of such plasma devices remain high, which can be decisive for their adoption in medical procedures. In this work, a simple modification of a low-cost plasma jet configuration is proposed in order to evaluate the device's viability for medical applications using argon as working gas. Typically, plasma jets operating with argon (Ar) generate high electrical currents, which makes them unsuitable for medical applications. With the proposed device modification, the treatment is no longer carried out using the plasma jet directly, but rather with the post-discharge effluent enriched with reactive oxygen and nitrogen species. As a result, the electrical current reaching the target remains below the safety threshold established for plasma applications in humans. Tests on water activation by plasma jets with and without the proposed modification indicate that both devices can yield comparable outcomes.
\end{abstract}

%
\\
\\
\\
\begin{tikzpicture}

\node[align=justify, text width=0.9*\textwidth, inner sep=1em]{
{\small This paper has been accepted for publication in Physica Scripta. This is the author's version which has not been fully edited and content may be different from the final publication. Citation information: F. do Nascimento, B. H. da Silva Leal, and K. G. Kostov, Phys. Scr., vol. 100, no. 9, p. 095601, Sep. 2025, DOI \href{https://dx.doi.org/10.1088/1402-4896/adfe2b}{10.1088/1402-4896/adfe2b}}
};

\node[xshift=3ex, yshift=-0.7ex, overlay, fill=white, draw=white, above
right] at (current bounding box.north west) {
\textit{Dear reader,}
};

\end{tikzpicture}

\section{Introduction}
Over the past two decades, plasma medicine has become a research field of great interest. Advances in equipment intended for medical applications have been quite significant. Currently, in several countries, there are commercially available plasma sources that, in some cases, are also certified for medical use by the local health regulatory agencies \cite{Duarte_comprehensive_2020, Bekeschus_medical_2021, Adesina_review_2024, Jeong_clinical_2024}. Most plasma sources employed in medical procedures produce cold atmospheric pressure plasmas (CAPPs). The beneficial effects of CAPP treatments are attributed to reactive oxygen and nitrogen species (RONS) generated within the plasma discharges \cite{Khlyustova_important_2019, Koga-Ito_cold_2024}.

Plasma can interact with biological targets through two primary means \cite{Busco_emerging_2020, Laroussi_cold_2020}. One of them is the direct plasma treatment, that is, with the plasma directly impinging on the target surface \cite{Busco_emerging_2020, Laroussi_cold_2020}. The other method is the indirect plasma treatment, where the plasma plume itself does not touch the target; however, the plasma-induced reactive species in the jet effluent interact with the surface \cite{Bekeschus_medical_2021, Laroussi_cold_2020}. Indirect plasma treatment, in turn, can also be performed using plasma-activated liquids (PAL) \cite{Koga-Ito_cold_2024, Laroussi_cold_2020, woedtke_foundations_2022}. The Direct method allows the simultaneous delivery of RONS, ultraviolet (UV) radiation, charged particles, and electric fields, which can induce effects like antimicrobial activity, wound healing, blood coagulation, etc. \cite{Laroussi_cold_2020, Weltmann_future_2019}. Indirect treatments through plasma effluent avoid direct electric field and charged particles effects but can still deliver long-lived RONS to biological samples \cite{Xiao_effects_2016}. Plasma-activated liquids, in turn, can transfer to the target only those RONS formed by the plasma jet that can be diluted in the treated liquid \cite{Montalbetti_production_2025, Wu_cold_2025}.

Previous research reported results on Direct plasma treatments in comparison with Indirect treatments using the post-discharge effluent, with both methods being suitable for medical and biomedical applications \cite{Thiyagarajan_characterization_2012, Attri_mechanism_2016, Saadati_comparison_2018, Lin_enhancement_2019}. Thiyagarajan \textit{et al} applied plasma jets in Direct and Indirect modes on antibiotic-resistant bacteria, with better results on bacteria inactivation obtained for the Direct condition \cite{Thiyagarajan_characterization_2012}. Attri \textit{et al} obtained very similar results on the degradation of dyes for both Direct and Indirect plasma treatments \cite{Attri_mechanism_2016}. Lin observed that the pH variation of physiological saline as a function of treatment time is almost the same for Direct and Indirect plasma exposure \cite{Lin_enhancement_2019}. On the other hand, Saadati \textit{et al} reported that Indirect treatment was not as effective as the Direct method in the treatment of melanoma cancer cells, but the former was found to be less cytotoxic to treated cells \cite{Saadati_comparison_2018}. 

When argon ($\rm{Ar}$) is used as the working gas, the plasma jets tend to exhibit filamentary patterns \cite{Wu_effects_2013, Xu_transition_2024}. This occurs because the discharge channel becomes highly conductive and very narrow at the same time. Consequently, the resulting plasma jet exhibits higher values of discharge current ($i_{dis}$), gas temperature ($T_{g}$), and, crucially for medical applications, leakage current, when compared to diffuse discharges. To avoid higher $T_{g}$ values when operating with $\rm{Ar}$ some plasma jet devices, employ high gas flow rates between {3 and 10 slm} \cite{reuter_kinpenreview_2018, Arndt_comparing_2018, yoshida_plasma_2020, Balazinski_safety_2025}. However, this solution has some disadvantages, such as {lower production} of RONS \cite{Hao_nitric_2014, He_transportation_2018, Srikar_development_2024}. Low $i_{dis}$ and leakage current values are usually achieved by increasing the distance between the plasma outlet and the target \cite{mann_introduction_2016, timmermann_piezoelectric-driven_2020}. However, this solution may not work for all cases, especially when $\rm{Ar}$ is chosen as the working gas \cite{Nastuta_cold_2022}.

{One of the most significant cost factors in plasma source development is the power supply. In our study, we employed a power supply adapted from a commercial aesthetic device, which significantly reduces the overall expenses. These devices are widely available on the consumer market and can be identified by searching online for ``high-frequency skincare''. The power supply used in this study was obtained from a device manufactured by Ibramed Ltda} \cite{noauthor_ibramed_nodate}{. Examples of similar devices from other manufacturers include the NuDerma from Pure Daily Care, USA }\cite{noauthor_nuderma_nodate}{ and the Darsonval from BactoSfera\textregistered, Ukraine} \cite{noauthor_darsonval_nodate}{. In a previous work from our research group, preliminary results of the development of a low-cost plasma source have been reported}\cite{do_nascimento_low_2024}{. Such a plasma source }is already capable of operating with $\rm{Ar}$ as the working gas. When configured for $\rm{Ar}$, safety parameters like gas temperature and UV radiation fell within established limits for medical applications. The production of ozone ($\rm{O_3}$) and nitrogen oxides ($\rm{NO_x}$) only slightly exceeded the safety limits {of 0.055 ppm for $\rm{O_3}$ and 0.019 ppm for $\rm{NO_2}$} \cite{noauthor_directive_2002, noauthor_directive_2008}. However, the leakage current values significantly surpassed 100 {\textmu}A, which is the threshold value established as safe for applications in human tissues \cite{noauthor_iec_2015}. In this work, we propose a simple modification of the above-mentioned device configuration, which aims to avoid direct contact between the target and the plasma jet, thus ensuring that only the RONS generated by the Ar plasma are used for biomedical applications. In this way, since there is no direct contact between the plasma and the target, the value {of the electrical current reaching the target} is expected to be negligible. 

\section{Experimental setup and methods}

\subsection{Plasma source and setup overview}
\begin{figure}[htb]
\centering
\includegraphics[width=0.9\textwidth]{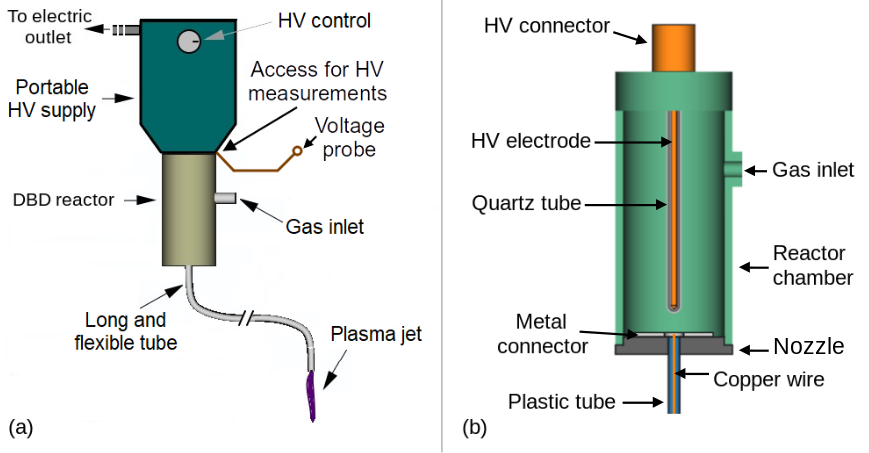}
\caption{Portable plasma source: (a) overview and (b) details of the DBD reactor. \label{plSrcSch}}
\end{figure}

Figure~\ref{plSrcSch} presents an overview of the plasma source (a) as well as a detailed view of the dielectric barrier discharge (DBD) reactor, which integrates the device (b). The plasma source comprises a portable power supply, attached to a dielectric barrier discharge (DBD) reactor, which is connected to a 1.0-meter-long flexible plastic tube with outer and inner diameters of 4.0 mm and 2.0 mm, respectively. The DBD reactor features a dielectric enclosure (inner diameter: 10 mm) housing a tungsten pin electrode (diameter: 1.8 mm), which is encased in a closed-end quartz tube with outer and inner diameters of 4.0 mm and 2.0 mm, respectively. The pin electrode is connected to a male metallic socket, which is mounted onto the dielectric enclosure and connects to the female socket of the power supply. Inside the flexible tube, a thin copper wire (diameter: 0.5 mm) runs along its length, anchored to a metallic connector located within the reactor chamber. {The copper wire inside the plastic tube acts as a floating electrode, and effectively transfers the plasma potential to the wires downstream tip. This action facilitates the ignition of the plasma jet right at the end of the long plastic tube }\cite{Kostov_study_2015, do_nascimento_comparison_2017}. The portable power supply, adaptation of a commercial aesthetic treatment device {(from Ibramed Ltda, Brazil} \cite{noauthor_ibramed_nodate}), generates damped sinusoidal waveforms with peak voltages reaching up to 20 kV and an oscillation frequency around 110 kHz. Each damped sine-wave function is a high-voltage (HV) burst signal (refer to Fig. 1(b) in \cite{nascimento_gas_2023} for a detailed depiction of the HV waveform). Besides the high-voltage oscillations within each burst, the power supply {can be adjusted to deliver from one to} four pulses per 50 or 60 Hz cycle, depending on the local power grid frequency. {The limitation on the number of HV-bursts has both advantageous and disadvantageous implications. The advantages pertain to controlling the gas temperature, leakage current, and UV-radiation emission. These parameters typically increase with the number of pulses, so by imposing a limitation, the device's operation is maintained within a safe range} \cite{do_nascimento_low_2024}{. Conversely, the disadvantageous implications include a reduction in the discharge power, a shorter effective discharge duration, and a reduced generation of reactive species in the plasma jet. In this work, the power supply was adjusted to deliver three pulses per cycle. The power grid in the laboratory operates at 60 Hz.}

\begin{figure}[htb]
\centering
\includegraphics[width=0.9\textwidth]{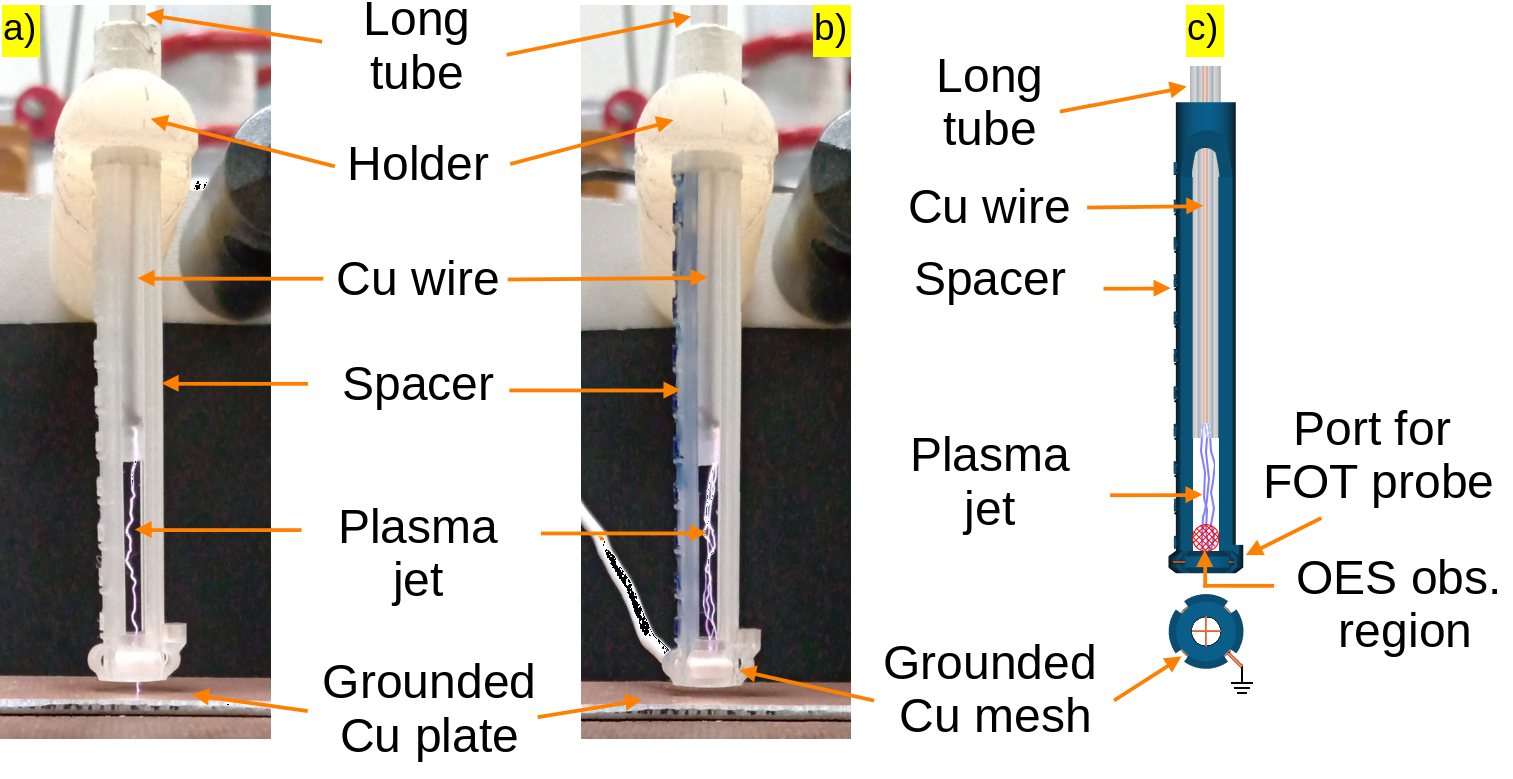}
\caption{(a) Direct and (b) Indirect setups for plasma treatments and (c) details of the spacer construction. The grounded $\rm{Cu}$ {mesh} is used only in the indirect treatment assembly. \label{setupDirInd}}
\end{figure}

In this work, we propose a simple modification of the plasma jet exit configuration that is intended to allow the device's medical applications when $\rm{Ar}$ is used as working gas. The modification consists of attaching to the end of the long plastic tube a 3D-printed spacer (to be referred to just as spacer from now on), {made of epoxy resin, as depicted in {Figures/Fig.~\ref{setupDirInd}}. The spacer has a cylindrical geometry with outer and inner diameters equal to 7.0 mm and 4.0 mm, respectively, with 70 mm in height. The spacer terminates with a grounded $\rm{Cu}$ mesh (2 crossed Cu wires),} which serves as a target to the plasma jet (see Fig. ~\ref{setupDirInd}). In this way, the electrical current produced by the plasma discharge does not reach the target. {Indeed, by comparing the photos in the Figs. 2(a) and 2(b), it is evident that the current filaments do not touch the metal target beneath the grounded Cu mesh.} However, the post-discharge effluent still contains some long-lived reactive species produced by the plasma that can be employed for medical purposes. This can then be called Indirect treatment. Some properties of plasma jets obtained when operating in Indirect mode were compared to those measured when the plasma jet impinges directly on a target (Direct treatment). The experimental setups for Direct and Indirect plasma treatments are depicted in Fig.~\ref{setupDirInd}. In both cases, the spacer was used to constrict the movement of the long tube in the horizontal plane and keep its displacement only along the vertical axis. Although the spacer is not strictly necessary for the Direct plasma jet configuration, it was still used (without the $\rm{Cu}$ mesh) to minimize possible differences between the two setups.

All parameters and properties of the plasma jets were investigated as a function of the distance from the plasma outlet and the target. In the case of Direct treatment, a grounded $\rm{Cu}$ plate was used as a target for the electrical characterization. For optical emission spectroscopy (OES) characterization, 10 ml of water samples, poured into a Petri dish, served as a target. The thermal characterization was performed for both targets, the $\rm{Cu}$ plate and the water sample. For the Indirect treatment configuration, the current was always measured on the grounded $\rm{Cu}$ mesh, but a $\rm{Cu}$ plate (or a water sample) was placed below the spacer as well. Both the $\rm{Cu}$ plate and the water samples were positioned at a distance of 1 mm from the bottom of the spacer in each case. 
For all experiments performed in this work, the gas flow rate was kept at 2.0 slm. 

\subsection{Electrical characterization}
The device's electrical characterization primarily involves obtaining the discharge power ($P_{dis}$) and effective discharge current ($i_{dis}$) as functions of the distance between the plasma outlet and the target, where the latter is the $\rm{Cu}$ plate in the Direct treatment condition or the $\rm{Cu}$ mesh in the Indirect case. For this purpose, the applied voltage ($V(t)$) was measured with a 1000:1 voltage probe (Tektronix, model P6015A), while for acquiring the current waveform, $i_{Targ}(t)$ we measured the voltage across a 47 ohms shunt resistor, placed after the target. When operating in the indirect condition, the electrical current collected at the grounded $\rm{Cu}$ plate ($i_{CuP}(t)$) was also measured. A 10.0 V/A current monitor (Pearson Electronics Inc., model 8600) was employed for this purpose, and these measurements were used to obtain the effective current reaching the $\rm{Cu}$ plate ($i_{Cu~plate}$). All the electrical signals were recorded with a four-channel oscilloscope (Rigol, model DS1104Z). The $P_{dis}$ values were then calculated using the recorded voltage and current signals as \cite{Pipa_equivalent_2019}:

\begin{equation}
{P_{dis} = \frac{1}{\tau} \int _{t_0} ^{t_0 + \tau} V(t) \cdot i_{Targ}(t) dt} \label{eq1}
\end{equation}

\noindent where $\tau$ = 16.67 ms.

{In order to improve the accuracy in the electrical measurements, the $V(t)$ and $i(t)$ waveforms were recorded for ten consecutive plasma discharges for each distance value between the plasma outlet and target. Thus, the final values of the electrical parameters presented in this work are the average values of ten consecutive measurements.}

\subsection{Thermal characterization}
Gas temperature ($T_{g}$) was measured using a fiber optic temperature (FOT) sensor (Weidmann Technologies Deutschland GmbH, Germany), with precision of 0.2 K. When measuring the temperature of the plasma jet, the FOT sensor was placed at the bottom of the spacer, 1 mm above the position of the $\rm{Cu}$ mesh (as indicated in Fig. ~\ref{setupDirInd}), with its tip always touching the plasma jet. The data acquisition for gas temperature measurements was performed as a function of the distance from the outlet to the target – water surface for Direct treatment configuration, or grounded $\rm{Cu}$ mesh for the Indirect one. {The FOT sensor was set to collect 25 temperature values in an interval of 0.5 s between consecutive measurements for each distance value.} Additionally, the temperature of the jet effluent alone was measured for the Indirect configuration. This was done by placing the FOT sensor tip right below the spacer.

\subsection{Optical emission spectroscopy characterization}
Optical emission spectroscopy was employed for the identification of emitting species and for measurements of the intensity of light emitted by selected species. The OES measurements were carried out with the plasma jet impinging on the water surface for the Direct configuration and on the $\rm{Cu}$ mesh for the Indirect one. The light emitted by the plasma jet was collected and guided to the spectrometer through optical fibers. In this work, the light emitted by the plasma was collected in the region close to the end of the spacer, 2 mm above the $\rm{Cu}$ mesh position, as it is highlighted in Fig. ~\ref{setupDirInd}. A multi-channel spectrometer from Avantes (model AvaSpec-ULS2048X64TEC), with an instrumental broadening (FWHM) equal to 0.76 \textpm~0.02 nm was employed to obtain an overview of the emission spectra in the wavelength range from 190 nm to 750 nm. Detailed OES measurements were performed with a multi-channel spectrometer from Horiba (model MicroHR), whose instrumental broadening is 0.34 \textpm~0.01 nm.

\subsection{Application in water functionalization}
In order to perform a semi-quantitative evaluation of the amount of reactive species that effectively reach the target upon Direct or Indirect plasma treatment, water samples were exposed to the plasma jet and effluent, respectively. Petri dishes filled with 10 ml of distilled water were placed under the spacer in a way that the distance between the bottom of the spacer and the water surface was nearly 1 mm. The water samples were then exposed to the plasma jet or to the effluent for 3 minutes.

After plasma exposure, the plasma-treated water (PTW) samples were analyzed using absorption spectroscopy. Absorption curves in the wavelength range from 200 nm to 280 nm were acquired on an UV–Vis spectrometer from Perkin Elmer (model Lambda25). {In order to enhance the reliability of the measurements, the analysis of the treated samples was conducted in the same sequence as their plasma exposure. This methodology was implemented to mitigate any potential variations that could arise between the time of exposure and the time of sample analysis.} The kinds of RONS formed in the liquid phase were then investigated by fitting the experimental absorbance curve with absorbance data for various RONS ($f_j = f_{RONS_{j}}$) as a function of the wavelength ($\lambda$). These absorbance data were extracted from works that presented absorbance as a function of the wavelength for RONS in aqueous medium \cite{Ling_uv-vis_2013, oh_-situ_2016, oh_uvvis_2017, Liu_quantifying_2019}. The equation used to perform the curve fitting has the following general form:

\begin{equation}
{g(\lambda) = \sum_{j} m_j f_j(\lambda)} \label{eq3}
\end{equation} 

\noindent where $m_j$ are multipliers of $f_j$($\lambda$) and g($\lambda$) is the resulting curve. If the UV-Vis system used to measure the absorption spectrum has absorbance calibration, then the $m_j$ multipliers return the concentrations of each of the RONS in the plasma-treated water \cite{oh_uvvis_2017, Petkovic_assessment_2024}.
\section{Results and discussion}

\subsection{Discharge power and current}
\begin{SCfigure}[1][htb]
\centering
\includegraphics[width=0.5\textwidth]{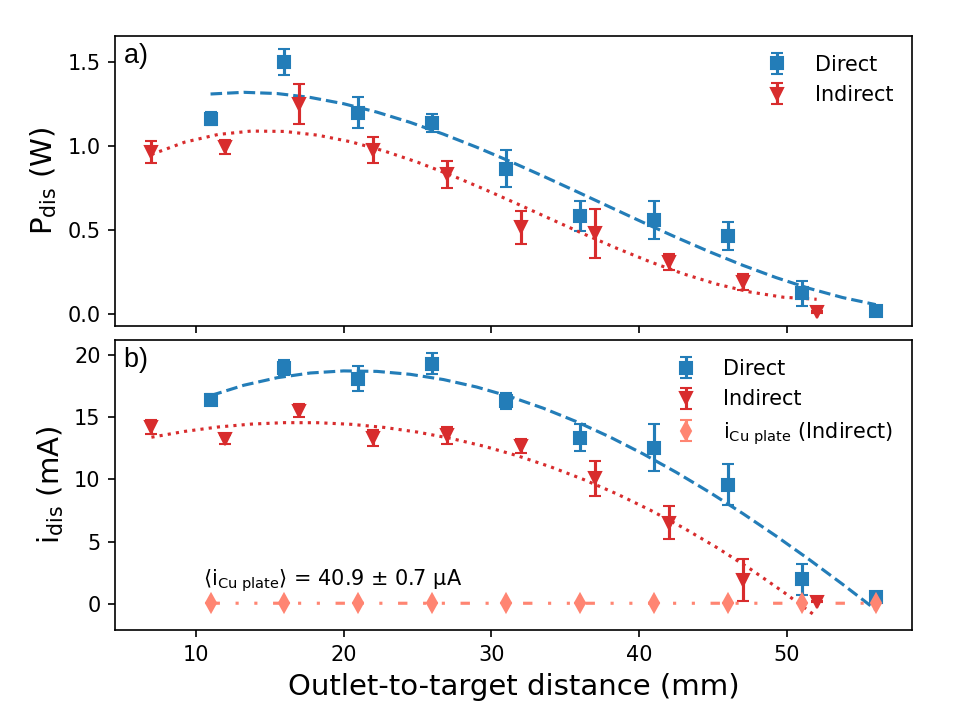}
\caption{(a) Discharge power and (b) effective discharge current as a function of the distance from the plasma outlet to the first metal target – $\rm{Cu}$ plate and $\rm{Cu}$ mesh for Direct and Indirect conditions, respectively. The curve of effective current at the $\rm{Cu}$ plate in the Indirect condition ($i_{Cu~plate}$) is also plotted in the graph. Dashed and dotted lines represent trend curves. \label{PIvsd}}
\end{SCfigure}

Figure~\ref{PIvsd} displays the discharge power ($P_{dis}$) and effective discharge current ($i_{dis}$) as functions of the distance from the plasma outlet to the target ($h$). The effective current collected at the $\rm{Cu}$ plate in the Indirect condition ($i_{Cu~plate}$) as well as its average value are also shown in the same graph.

{When comparing the curves obtained in the case of Direct and Indirect process in the {Figures/Fig.~\ref{PIvsd}} it can be noticed that the values of both, $P_{dis}$ and $i_{dis}$ for the Indirect treatment, are always slightly smaller than the corresponding ones for the Direct process. Such discrepancies are possibly related to different charge accumulation on the insulating spacer in the two cases. Since in the Indirect case the current is collected by a grounded mesh inside the spacer, the charge accumulation on the spacer's inner surface is higher than in the Direct treatment case, in which the plasma jet directly touches the grounded Cu plate.}

The current $i_{Cu~plate}$ that flows to the Cu plate in the Indirect condition as a function of the outlet-to-target distance is almost the constant with an average value of 40.9 \textpm~0.7 {\textmu}A. The latter is well below the value established as safe for devices aimed for medical applications (100 {\textmu}A). The $i_{Cu~plate}$ current probably originates from some charged particles, not collected by the $\rm{Cu}$ mesh, that are carried by the effluent and reach the target.

\subsection{Gas temperature measurements}
\begin{SCfigure}[1][htb]
\centering
\includegraphics[width=0.5\textwidth]{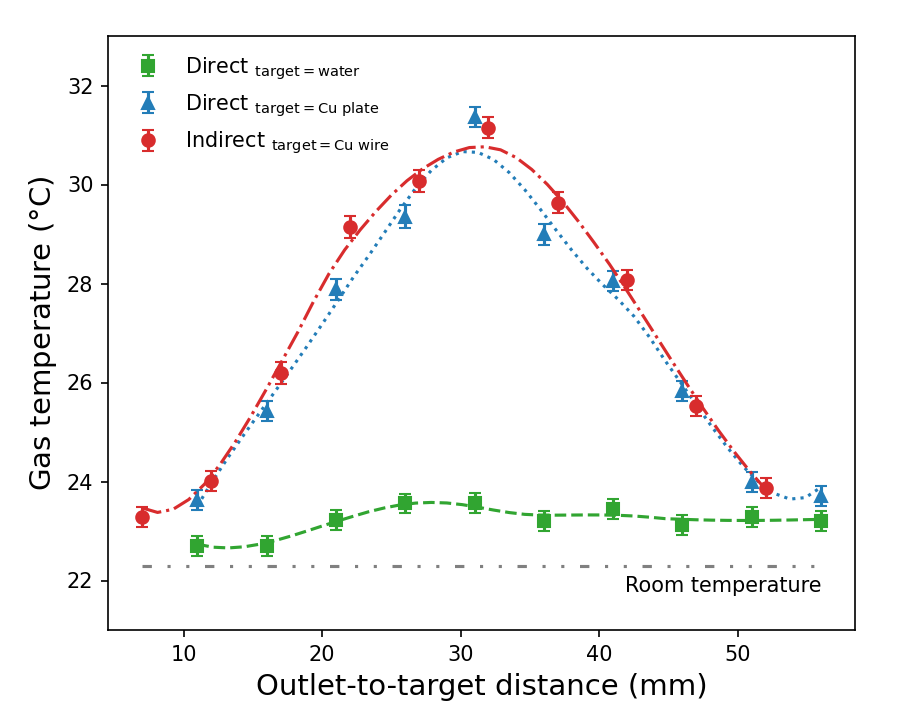}
\caption{Gas temperature measurements as a function of the outlet-to-target distance for Direct and Indirect conditions. In the Direct case, $T_{g}$ was measured with the plasma jet impinging on both the water surface (green squares) and $\rm{Cu}$ plate (blue triangles). In the Indirect case, the $T_{g}$ was measured with the plasma jet impinging on the $\rm{Cu}$ mesh in the spacer. Dashed and dotted lines represent trend curves. \label{gasTemp}}
\end{SCfigure}

The measurements of $T_{g}$ as a function of the distance between the plasma outlet and the target are presented in Fig.~\ref{gasTemp}. As it can be seen, the $T_{g}$ values measured with the plasma jet impinging on the water surface (Direct case) are very close to the room temperature, presenting small variations (less than 1 {\textdegree}C) as the distance from the plasma outlet increases. When operating with the plasma jet impinging on metal targets, in both Direct and Indirect conditions, there are only small differences in the $T_{g}$ curves as a function of the outlet-to-target distance. In both cases, the $T_{g}$ values start to increase as $h$ increases, reach a peak temperature near $h =$ 30 mm, and then decrease as $h$~increases. {This temperature behavior is likely due to the Ohmic heating of the plasma jet, since the discharge current is the main contributor to the gas heating in atmospheric pressure plasma jets} \cite{Xian_discharge_2016, Slikboer_revealing_2020}{. As it can be seen in {Figures/Fig.~\ref{PIvsd}}, the measured $i_{dis}$ exhibits minimal variation for $h$ up to 25 mm. Consequently, with increasing $h$, the segment of the plasma jet heated by $i_{dis}$ also elongates. This leads to an elevation in the gas temperature at the measurement point, which is moved further from the plasma outlet as $h$ increases. Beyond this initial range, the larger distance between the plasma outlet and the target leads to a substantial reduction in $i_{dis}$. Therefore, the thermal contribution of $i_{dis}$ to the bulk gas temperature becomes less pronounced, particularly when compared to the cooling effects arising from the gas flow and interaction of the plasma jet with the environment.}

The $T_{g}$ values with the plasma jets impinging on water or $\rm{Cu}$ surfaces differ significantly likely because of the large difference in the discharge {current} achieved for targets with very different electrical conductivity values{, typically 5.8$ \times 10^7$ S/m for copper and 1.0-2.0$ \times 10^{-4}$ S/m for distilled water} \cite{Matula_electrical_1979, Ageev_features_2020}.

In any case, the $T_{g}$ values remain significantly lower than the 40 {\textdegree}C threshold, which is the gas temperature limit recommended for medical applications of plasma jets \cite{mann_introduction_2016}.

\subsection{Identification of emitting species}
\begin{SCfigure}[1][htb]
\centering
\includegraphics[width=0.7\textwidth]{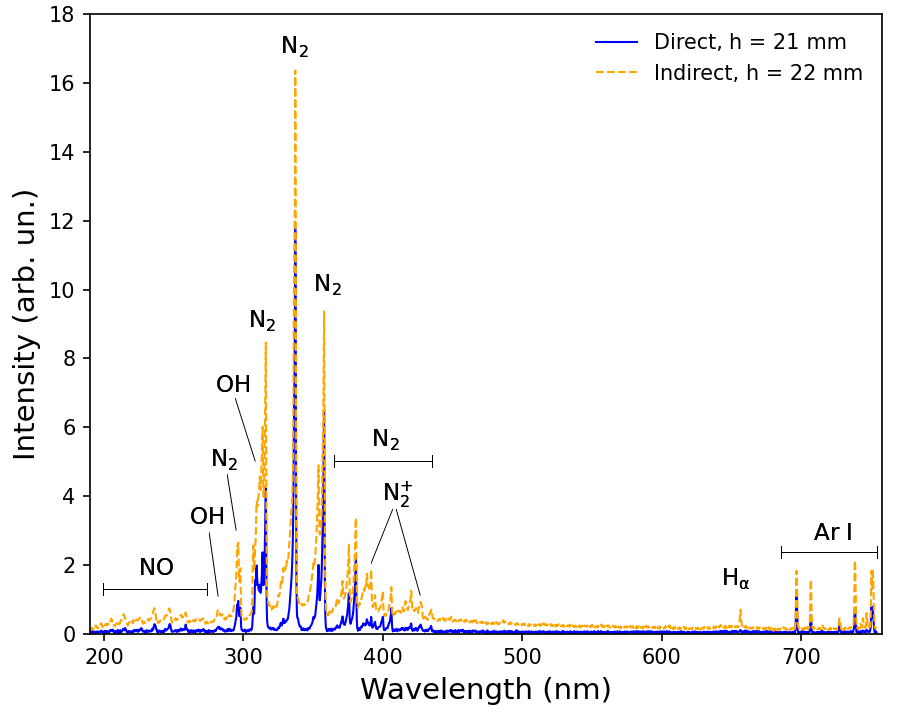}
\caption{Typical emission spectra of the plasma jets for Direct and Indirect configurations. \label{typSpec}}
\end{SCfigure}

Figure~\ref{typSpec} displays typical OES spectra measured for the Direct and Indirect configurations. For the Direct case, the spectrum was recorded with the plasma jet impinging on the water surface for an outlet-to-target distance ($h$) equal to 21 mm. In the Indirect case, the plasma jet was impinging on the $\rm{Cu}$ mesh. Nevertheless, the water sample was kept below the spacer. Thus, the OES measurements were carried out in the same configuration as it was in the assays for applications in water functionalization.

From Fig.~\ref{typSpec} it can be seen that the excited species in the plasma jet are almost the same, except for the hydrogen emission ($\rm{H_\alpha}$) observed at 656.28 nm. In both cases, the excited molecular species found in the spectra were $\rm{NO}$, $\rm{OH}$, $\rm{N_2}$ and $\rm{N_2^{+}}$. Line emissions of neutral argon atoms ($\rm{Ar~I}$) are also present in both spectra, as well as line emissions from neutral oxygen atoms ($\rm{O~I}$, not shown in Fig.~\ref{typSpec}, measured with the Horiba spectrometer).

Another visible difference between the spectra recorded for Direct and Indirect configurations is the presence of a continuum emission in the Indirect spectrum, which lifts up the spectrum intensity in almost the whole wavelength range shown in Fig.~\ref{typSpec}. This continuum emission is likely due to bremsstrahlung radiation and is mostly linked to collisions among electrons and neutrals in the plasma \cite{Park_continuum_2015, Nikiforov_electron_2015}. It is an indication of higher electron density and temperature values for the plasma jet impinging on the grounded $\rm{Cu}$ mesh, when compared to it impinging on the water \cite{Park_continuum_2015, Nikiforov_electron_2015}. {The electron density and temperature of the plasma jet are both influenced by the discharge power, which is a function of the applied voltage and discharge current } \cite{Nikiforov_electron_2015, Xiao_electron_2014, Lin_average_2018, Giuliani_relation_2020}. {Given that the applied voltage is constant across both operating conditions of {Figures/Fig.~\ref{typSpec}}, the change in discharge power is solely a result of the variation in discharge current. Since water has a lower conductivity than the Cu wire, the discharge current is higher in the Indirect case, leading to increased electron density and temperature when compared to the Direct condition.}

Although the OES measurements were performed using the same arrangement for both Direct and Indirect conditions, the intensity of the light emitted by the reactive species can not be directly compared because the OES measurements were made at different positions in the plasma column in relation to the target. That is, in the Direct configuration, the OES measurements were performed {$\sim$}7 mm above the water surface, while in the Indirect condition, the measurements were carried out {$\sim$}3 mm above the $\rm{Cu}$ mesh. In addition, the continuum emission observed in the spectra measured for the Indirect configuration makes the intensity of the reactive species look higher than they actually are. However, a qualitative comparison of the intensities as a function of the outlet-to-target distance is still valid and is therefore more relevant for the purposes of this work.

\begin{SCfigure}[1][htb]
\centering
\includegraphics[width=0.5\textwidth]{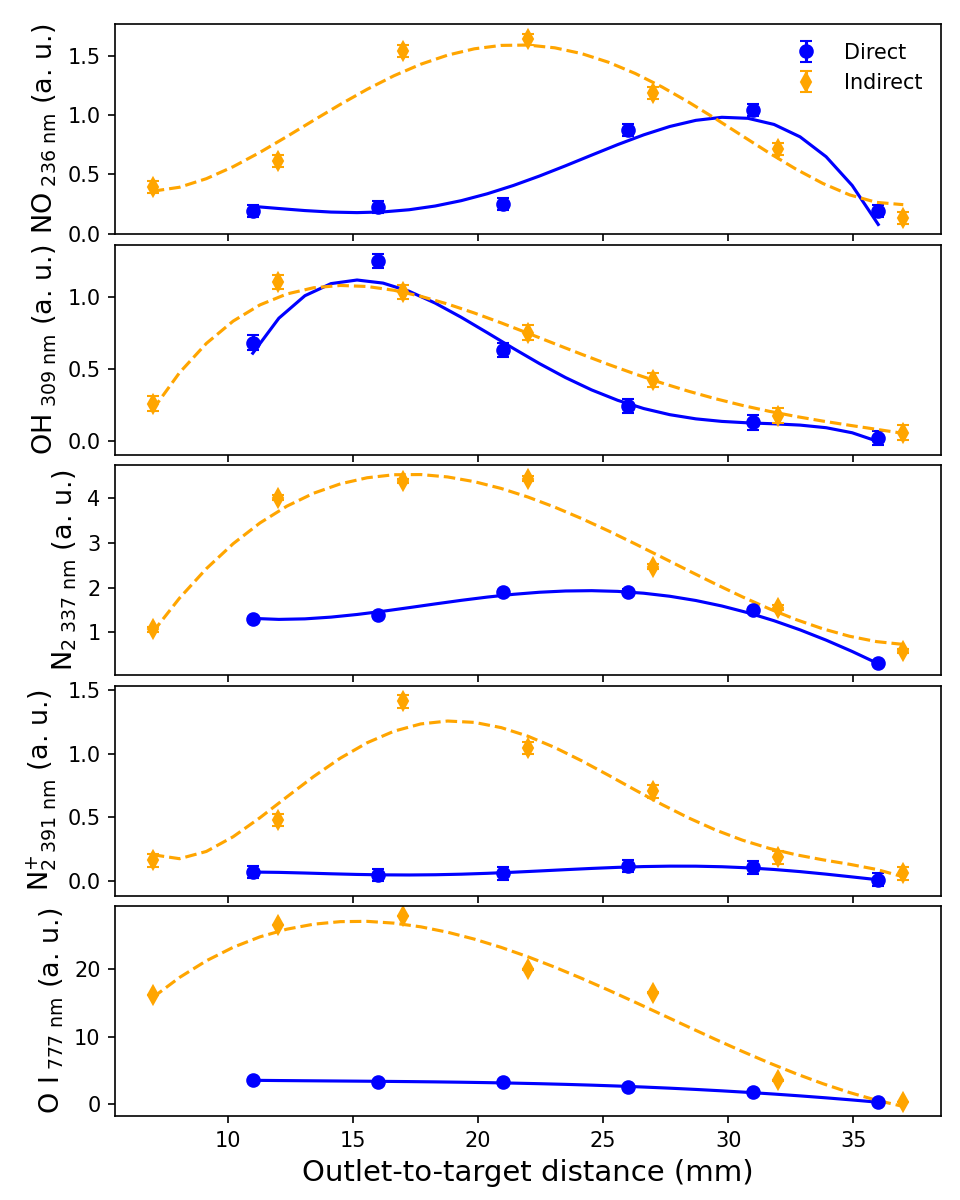}
\caption{Intensities of admitting species as a function of the distance between the plasma outlet and target for selected reactive species. \label{intVsh}}
\end{SCfigure}

Figure~\ref{intVsh} shows the intensities of the light emitted by selected species as a function of the distance from the plasma outlet and target (water surface and $\rm{Cu}$ mesh, for Direct and Indirect conditions, respectively). All measurements presented in Fig.~\ref{intVsh} were done using the Horiba spectrometer. From Fig.~\ref{intVsh} it can be clearly seen that the behavior of the intensity as a function of $h$ differs for almost all curves when comparing Direct and Indirect configurations. {It is probable that the alteration of relative air humidity in the vicinity of the plasma jet, induced by the water target in the Direct case, leads to a change in the axial distribution of the reactive species, with the exception of $\rm{OH}$, whose curves have a similar trend in both configurations,} presenting peak values at $h$ close to 15 mm.

Although the absolute intensity values measured under different conditions can not be directly compared, it is worth mentioning that the intensity of the $\rm{O~I}$ line emissions measured for the Indirect case is much higher than those observed for the Direct condition. Since in the wavelength range where the $\rm{O~I}$ emissions are measured, the continuum emission from bremsstrahlung radiation is almost negligible, the differences in intensities are then due to the higher density of oxygen atoms in an excited state. Therefore, it can be inferred that the production of oxygen atoms in the Indirect configuration is considerably higher than in the Direct one.

\subsection{Water treatment results}
\begin{figure}[htb]
\centering
\begin{minipage}{0.49\textwidth}
\centering
\includegraphics[width=\textwidth]{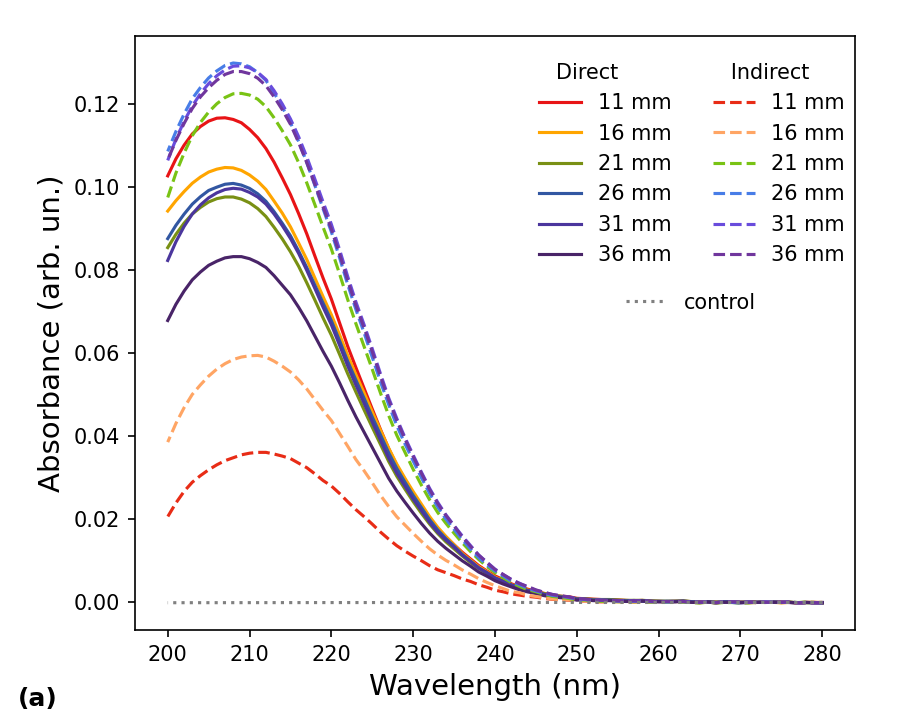}
\end{minipage}
\begin{minipage}{0.49\textwidth}
\centering
\includegraphics[width=\textwidth]{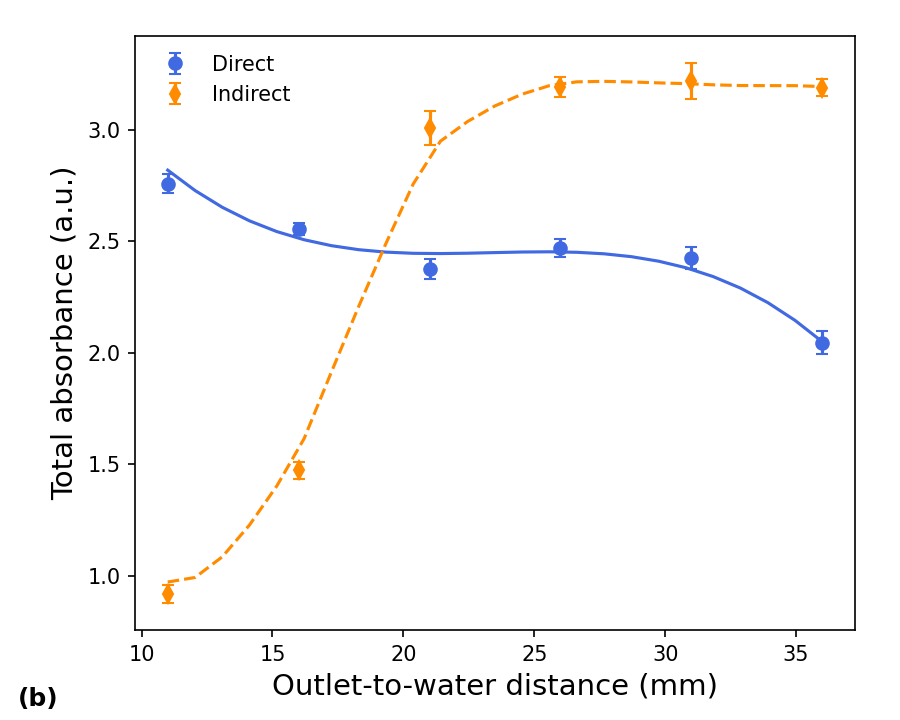}
\end{minipage}
\caption{(a) Absorption spectra of water samples exposed to Direct and Indirect plasma treatment for various distances from the plasma outlet to the water surface ($h_w$). (b) Total absorbance as a function of $h_w$. In all experiments, the volume of liquid and the plasma exposure time were 10 ml and 3 minutes, respectively. \label{uvVizAll}}
\end{figure}

In this section, the results obtained when water samples were exposed to Direct and Indirect plasma treatment are presented. The results of UV-Vis analyses of plasma-treated water (PTW) samples are shown in Fig.~\ref{uvVizAll}. All the samples in Fig.~\ref{uvVizAll} were exposed to plasma for 3 minutes. In Fig.~\ref{uvVizAll}(a), the absorption spectra of all treated samples are presented. In Fig.~\ref{uvVizAll}(b), the total absorbance ($A_{Tot}$ = area under the curve for each sample in Fig.~\ref{uvVizAll}(a) is plotted as a function of the outlet to water distance ($h_w$), comparing the Direct and Indirect operating conditions. The total absorbance of a water sample is related to the amount of RONS retained by the sample after plasma exposure. From Fig.~\ref{uvVizAll}(b) it can be seen that the $A_{Tot}$ values are higher in the Direct case for shorter $h_w$ values ($h_w < $ 20 mm). However, as $h_w$ increases, the $A_{Tot}$ values for the Indirect configuration become higher than the ones for the Direct case, reaching a plateau after $h_w > $ 20 mm. {Referencing {Figures/Fig.~\ref{intVsh}}, It is evident that the Indirect plasma configuration does not yield a plateau in the production of gas-phase reactive species for axial distances $h_w > $ 20 mm, indicating a variation in species generation beyond this point. This apparent discrepancy occurs because the OES measurements were spatially constrained, permitting analysis only within a limited region of the plasma jet. Conversely, the water samples exposed to plasma treatment are capable of accumulating and retaining reactive species generated along the entire plasma column. This kind of ``accumulation effect'' likely results in the saturation of RONS absorbed within the treated aqueous medium, thereby accounting for the observed differences. Regarding the direct configuration, $A_{Tot}$ presents a decreasing trend as $h_w$ increases.}

\begin{figure}[htb]
\centering
\begin{minipage}{0.49\textwidth}
\centering
\includegraphics[width=0.9\textwidth]{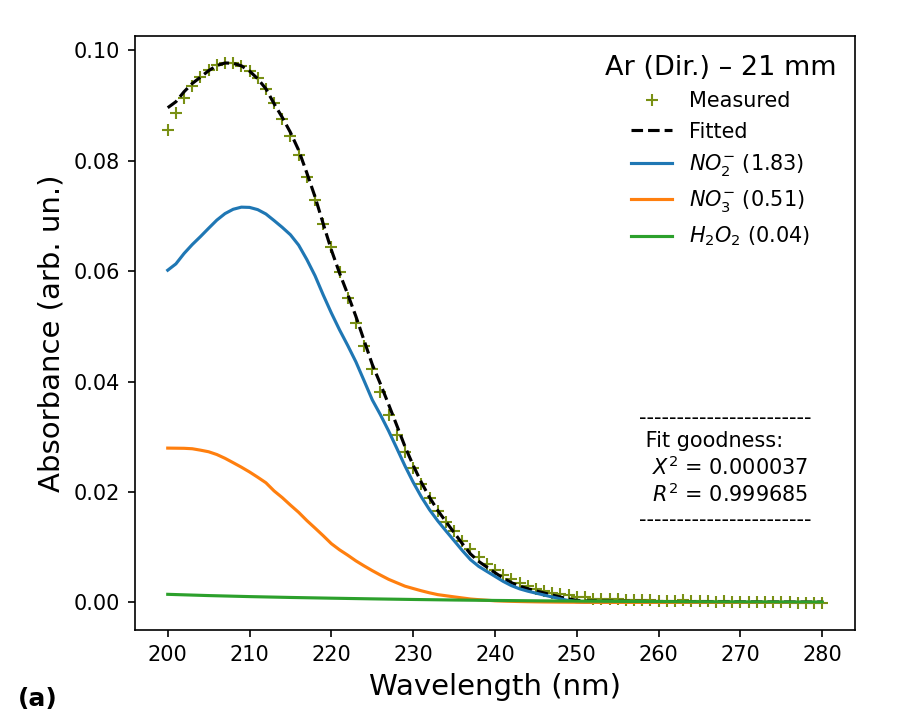}
\end{minipage}
\begin{minipage}{0.49\textwidth}
\centering
\includegraphics[width=0.9\textwidth]{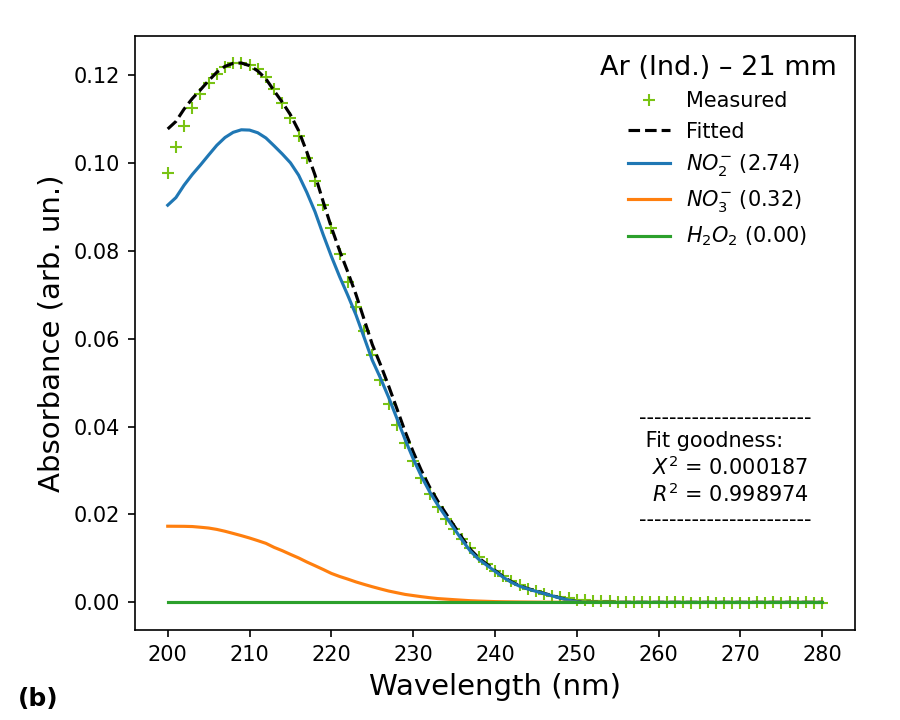}
\end{minipage}
\caption{Example of the fitting procedure for RONS quantification in water for (a) Direct and (b) Indirect plasma exposure. The numbers in parentheses are the values of the area under the curve of the corresponding species. \label{speciesFitEgizAll}}
\end{figure}

\begin{SCfigure}[1][htb]
\centering
\includegraphics[width=0.5\textwidth]{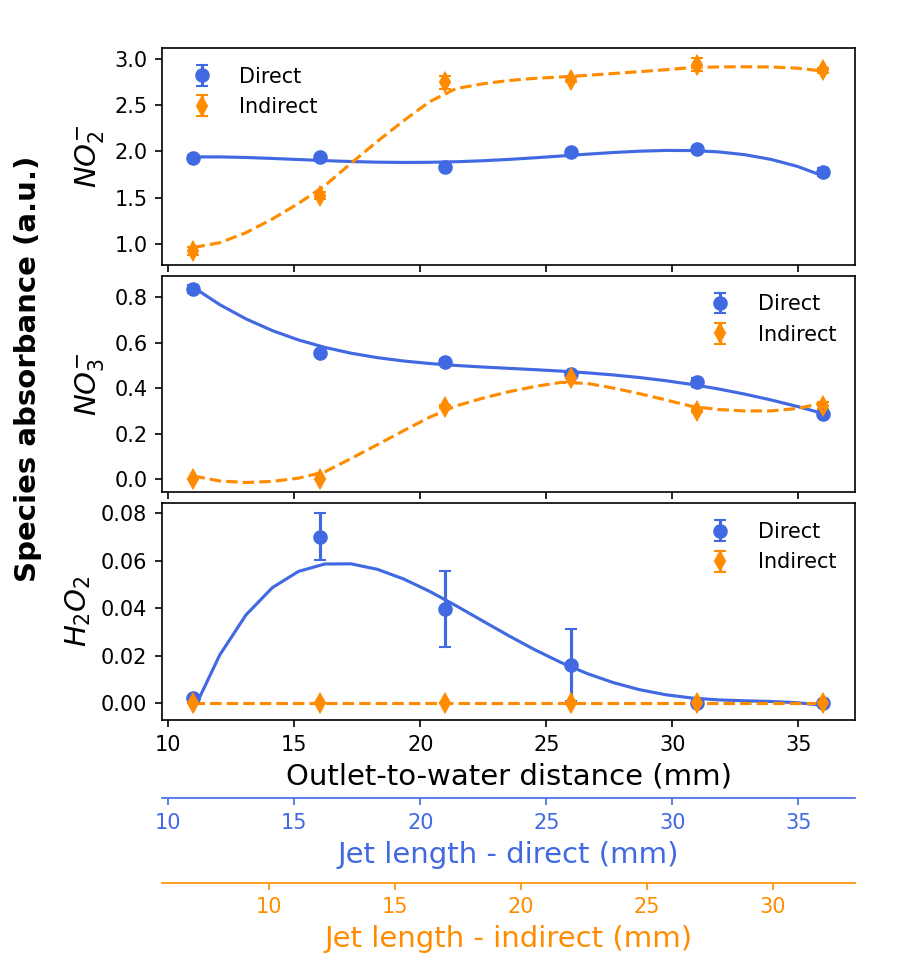}
\caption{Total absorbance of the detected reactive species as a function of the distance between plasma outlet and water surface. The plasma jet lengths in the Direct and Indirect exposure cases are also indicated in the figure. \label{speciesAsorb}}
\end{SCfigure}

A more detailed analysis of the absorbance spectra shown in Fig.~\ref{uvVizAll}(a) was performed for each curve to identify the different RONS generated in the PTW. In Fig.~\ref{speciesFitEgizAll} are presented examples of the fitting process for identification of the RONS in the PTW for (a) Direct and (b) Indirect treatment. {The numbers in parentheses shown in  {Figures/Fig.~\ref{speciesFitEgizAll}} are the values of the area under the absorption curve for each RONS identified in the fitting procedure. These values also correspond to the absorbance by each species ($A_{RONS}$).} Figure~\ref{speciesAsorb}, for instance, presents curves of $A_{RONS}$ as a function of $h_w$ for the RONS detected in the PTW samples. The RONS species detected in the liquid phase were nitrite ($\rm{NO_2^{-}}$), nitrate ($\rm{NO_3^{-}}$), and, when operating with the Direct configuration, hydrogen peroxide ($\rm{H_2O_2}$). From Fig.~\ref{speciesAsorb} it can be seen that $\rm{NO_2^{-}}$ is the most abundant species detected in the PTW samples, for both operating conditions, followed by $\rm{NO_3^{-}}$. A small amount of $\rm{H_2O_2}$ was detected only in some water samples exposed to the Direct treatment.

Regarding the operation in the Direct configuration, from Fig.~\ref{speciesAsorb} it can be seen that the amount of $\rm{NO_2^{-}}$ in the water samples practically does not vary for different $h_w$ values, which probably means that $\rm{NO_2}$ molecules are absorbed by the water at almost the same rate with which they are produced in the gaseous phase. On the other hand, the amount of $\rm{NO_3^{-}}$ in the water samples presents a clear tendency to decrease as $h_w$ increases. The $\rm{H_2O_2}$, for instance, presents a peak value at $h_w$ equal to 16 mm. However, the amount of $\rm{H_2O_2}$ detected in the liquid phase is very low and is probably within the detection limit of the curve fitting performed to identify RONS. This low production of $\rm{H_2O_2}$ is likely related to the low exposure time (3 min) chosen in this set of experiments.

Regarding the operation in the Indirect condition, the amount of $\rm{NO_2^{-}}$ detected in the water samples increases relatively fast with $h_w$ up to 21 mm. Beyond this value, the amount of $\rm{NO_2^{-}}$ remains nearly constant. Concerning the $\rm{NO_3^{-}}$ molecules, they were not detected in the water samples for $h_w \leq$ 16 mm. After that, there is an increase in its production, with a peak value at $h_w$ = 26 mm,  followed by a small reduction to lower values for larger distances of exposure. 

\begin{figure}[htb]
\centering
\begin{minipage}{0.49\textwidth}
\centering
\includegraphics[width=0.9\textwidth]{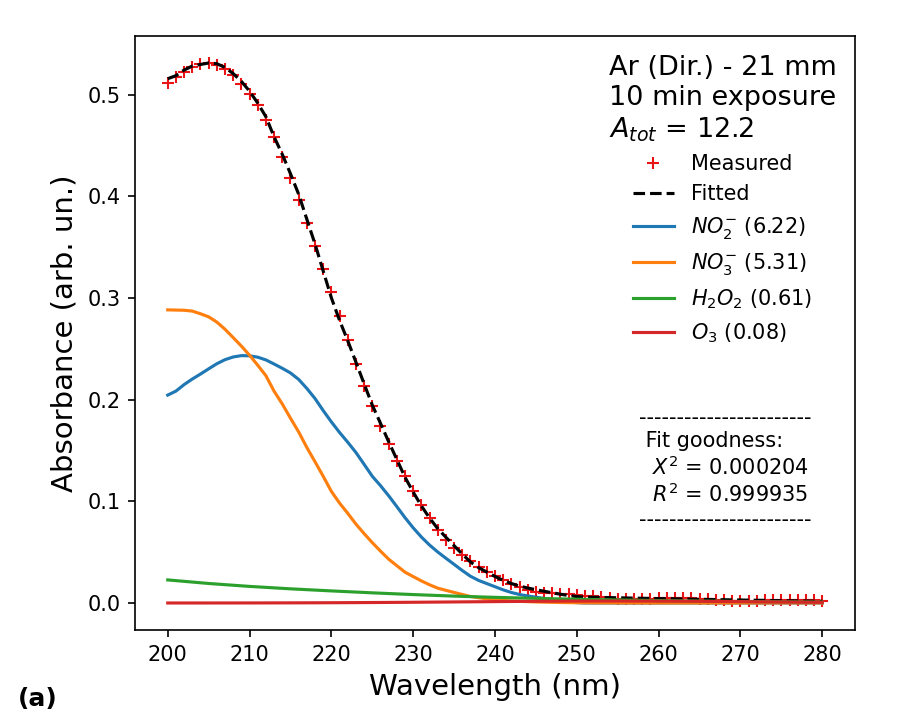}
\end{minipage}
\begin{minipage}{0.49\textwidth}
\centering
\includegraphics[width=0.9\textwidth]{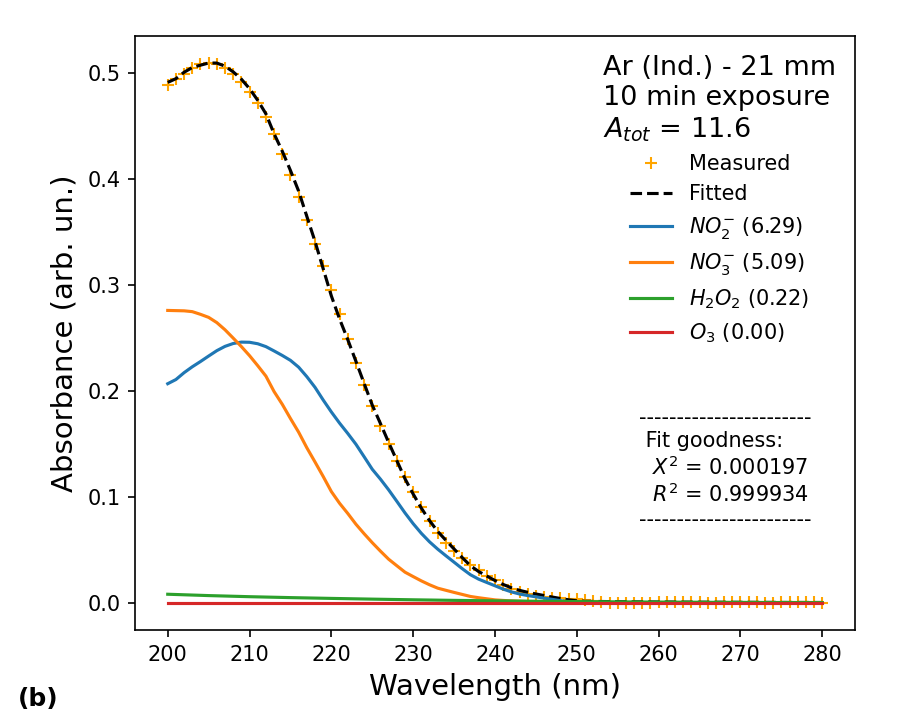}
\end{minipage}
\caption{Quantification of RONS in water for (a) Direct and (b) Indirect treatments for 10 minutes of plasma exposure time. The volume of the water samples was 10 ml in both cases. The numbers in parentheses are the values of the area under the curve of the corresponding species. \label{sFit10m}}
\end{figure}

In Fig.~\ref{sFit10m} are presented the results of RONS quantification in water samples exposed to (a) Direct and (b) Indirect treatments for 10 minutes at a distance of 21 mm between the plasma outlet and the water surface. { At 21 mm, a kind of balanced condition between both processes is achieved, with Indirect treatment generating high levels of long-lived $\rm{NO_2^{-}}$ and $\rm{NO_3^{-}}$ species, while there is still a relatively high production of $\rm{H_2O_2}$ in the Direct treatment. Thus, at this distance, the differences and complementarities between the modes of operation are most evident, which makes 21 mm the best choice to compare the overall production of reactive species using Direct and Indirect treatments.} For 21 mm distance and 10 min exposure time, the total absorbance for the Direct condition is nearly 5\% higher than the one obtained for the Indirect case. {Notably, for both processes using a treatment time of 10 min, the generated amount of $\rm{NO_3^{-}}$ was much higher,  and at some distances comparable or even exceeding the amount of $\rm{NO_2^{-}}$}.  It can also be seen in Fig.~\ref{sFit10m} that the production of $\rm{NO_2^{-}}$ in the water samples was slightly higher than that for the Indirect exposure and that the production of $\rm{NO_3^{-}}$ was higher for the Direct treatment. The amount of $\rm{H_2O_2}$ found in the treated water samples was three times higher using the Direct configuration compared to the Indirect one. Regarding the production of RONS, these data confirm the trend presented in Fig.~\ref{speciesAsorb} for $h_w$ = 21 mm. In addition, only in the case of Direct treatment of water, a small amount of ozone ($\rm{O_3}$) was detected.

A noticeable result shown in Fig.~\ref{speciesAsorb} is that the amount of $\rm{H_2O_2}$ detected in the liquid phase for the Indirect configuration is virtually zero. Considering that the $\rm{OH}$ production in the gaseous phase (see OES results in Fig.~\ref{intVsh}) is comparable in both operating conditions (with the caveats already discussed), it would be expected to find $\rm{H_2O_2}$ in the liquid phase in both conditions (due to the reaction $ OH + OH \rightarrow H_2O_2$), but this did not happen. Since $\rm{H_2O_2}$ was detected in the water samples treated for 10 minutes using both configurations of plasma exposure, this is an indication that just 3 minutes of plasma exposure is not enough for the production of $\rm{H_2O_2}$ in the liquid phase for the Indirect treatment. These results suggest that the production of $\rm{H_2O_2}$ occurs preferably in the aqueous phase rather than in the gaseous one. Although this may be a limitation for applications of the Indirect configuration, this fact can be used to selectively produce reactive species in situations where the presence of $\rm{H_2O_2}$ is undesirable.

{It is important to mention that the low values of the uncertainties in the electrical, optical and thermal measurements suggests that the plasma parameters are highly reproducible. This reproducibility, combined with the strict control of exposure time, contributes to a good reproducibility of the results obtained for water activation.}

{Considering the parameters under variation in this study, a general conclusion comparing the Direct and Indirect operating conditions is that both processes can be employed, depending on the operating conditions. Regarding the production of RONS in the liquid phase, this study demonstrated that for shorter distances between the plasma outlet and the water surface it is better to employ the Direct configuration, while for larger distances, the Indirect configuration is expected to yield better results. However, considering biological applications, the Indirect mode of operation is the best choice because it generates a leakage current below the threshold value for plasma equipment in medicine.}

\section{Conclusions}
This work aimed to verify whether a small design modification of the previously developed plasma jet device could make its operation with argon gas safe and potentially useful for biomedical applications. The modification in question includes the usage of a plastic spacer with a grounded copper mesh in it that is introduced at the plasma jet exit. The introduction of this spacer changed the way in which the plasma jet interacts with the target, making the treatment indirect. Therefore, to prove the effectiveness of this modification, a comparison with the direct plasma application was necessary.
Results from the electrical characterization demonstrated that the discharge power and effective current that reaches the first metal target ($\rm{Cu}$ plate and $\rm{Cu}$ mesh for Direct and Indirect cases, respectively), are practically equivalent. However, in the Indirect case, with a grounded copper plate acting as {a target, resembling the case of a biological object}, the measured electric current is much lower than the limit value considered safe for medical applications of plasmas. The results from the thermal characterization showed that in all cases the gas temperature is well below the safety level (i.e., 40 {\textdegree}C). 

Regarding the production of reactive species, the OES reveals that for both modes of device operation, the generation of RONS in the gaseous phase can be considered equivalent for almost all emitting species. An exception is the production of free oxygen atoms, which is considerably higher in the Indirect case. For the species detected in the aqueous phase, it was found that at higher distances between the outlet and the target, it is possible to generate more $\rm{NO_2^{-}}$ with the Indirect configuration than with the Direct mode of operation, maintaining an equivalent production of $\rm{NO_3^{-}}$. On the other hand, the production of $\rm{H_2O_2}$ during Indirect applications is usually much lower than the one that can be achieved with the Direct configuration.

{There are several important safety considerations for using plasma jets in medical applications. The gas temperature and leakage current must stay within safety limits to avoid damaging cells. Other factors like harmful gases and UV radiation should also be kept below recommended thresholds, as exceeding them can limit how long a patient can be exposed to the treatment. Beyond the patient, the safety of the operator is also crucial. The production of harmful gases must be controlled to ensure the operator is not exposed to unhealthy working conditions over many hours. While overall electrical safety and electromagnetic compatibility are important and should be addressed during the prototype phase, the plasma source discussed in the text already meets electrical safety standards because it uses a power supply from a product already certified by regulatory agencies in Brazil and other countries. This is an advantage, but it is noted that all parts of the device must be safe, not just the power supply.} As a general conclusion, from the electrical and thermal characterization results, it can be stated that the Indirect configuration offers safe operating conditions, whereas the results from RONS measurements, in both gaseous and liquid phases, indicate that the Indirect application produces the essential elements required for medical applications.

Further studies aimed at the optimization of reactive species production must be carried out for different gas flow rates as well as for extended exposure times. In addition, the application of the effluent produced by the Indirect configuration should also be tested in \textit{in-vitro} experiments to ensure its antimicrobial efficacy.

\section*{Data Availability Statement}
The data are contained in this manuscript. Raw data are available from the authors upon reasonable request.

\section*{Conflict of Interest}
The authors declare that there is no conflict of interest in this work.

\section*{Acknowledgments}
The authors thank Prof. Dr. Erick Siqueira Guidi for the discussion on the manufacturing of the 3D printed part. This work received financial support from the São Paulo Research Foundation (FAPESP), under Grants \#2019/05856-7 and \#2020/09481-5, and from Coordination of Superior Level Staff Improvement (CAPES), under Grant \#88887.912282/2023-00.

\section*{References}
\bibliographystyle{unsrt}
\bibliography{indirect_plasma_treatment}

\end{document}